\documentstyle[12pt]{article}
\def\thebibliography#1{\section*{
\rm\normalsize\hspace{-.2cm}{\bf4. References} }\list
 {[\arabic{enumi}]}{\settowidth\labelwidth{#1}\leftmargin\labelwidth
 \advance\leftmargin\labelsep
 \usecounter{enumi}}
 \def\newblock{\hskip .11em plus .33em minus .07em}
 \sloppy\clubpenalty4000\widowpenalty4000
 \sfcode`\.=1000\relax}

\renewcommand{\thefootnote}{\fnsymbol{footnote}}
\begin{document}
\begin{center}
\begin{Large}
\renewcommand{\thefootnote}{\fnsymbol{footnote}}
{\bf Exact $m_{quark}\neq 0$ Condensates in
$QCD_{1+1}(N_C\rightarrow \infty)$}
\end{Large}
\footnote{Talk presented at the `Workshop
on Quantum Infrared Physics', Paris, June 1994.}
\\[1.5ex]
M.Burkardt\\
Institute for Nuclear Theory,
Univ. of Washington, Seattle, WA 98195\\[2.ex]
\renewcommand{\thefootnote}{\arabic{footnote}}
\baselineskip8pt
\parbox[t]{13.9cm}{\small In the limit of an infinite number of colors,
we derive an analytic expression for the quark condensate
in $QCD_{1+1}$ as a function of the quark mass and the
gauge coupling constant. For zero quark mass, a nonvanishing
quark condensate is obtained. Nevertheless, we prove
that there is no phase transition as a function of the
quark mass. It is furthermore shown that the expansion
of $\langle 0 | \overline{\psi}\psi |0\rangle$ in the gauge
coupling
has zero radius of convergence but that the perturbation
series is Borel summable with finite radius of convergence.
The nonanalytic behavior
$\langle 0 | \overline{\psi}\psi |0\rangle
\stackrel{m_q\rightarrow0}{\sim} - N_C \sqrt{G^2}$ can only be
obtained by summing the perturbation series to infinite order.
}\\[2.ex]
\baselineskip12pt
\end{center}
\noindent
{\bf 1. Introduction}\\[1.5ex]
What is interesting about the quark condensate
in $QCD_{1+1}$? Zhitnitsky \cite{zhit:vac}, using QCD-sum rule
techniques, derived an exact result for the
condensate in the limit of an infinite number
of colors\footnote{Note that Coleman's theorem \cite{co:no}
prohibits spontaneous breakdown of chiral
symmetry for any finite number of colors, since this
is a $1+1$-dimensional model.}
and in the limit $m_q\rightarrow 0$
\begin{equation}
\left.\langle 0 | \overline{\psi}\psi |0\rangle
\right|_{m_q=0} =
- \frac{N_C}{\sqrt{12}}\sqrt{\frac{g^2C_F}{\pi}},
\label{eq:mis0}
\end{equation}
where $C_F=\frac{N_C^2-1}{2N_C}$ and $G^2 \equiv \frac{g^2C_F}{\pi}$
is held fixed as $N_C\rightarrow \infty$.
This result is remarkable in several respects:
Firstly, $\left. \langle 0 | \overline{\psi}\psi |0\rangle
\right|_{m_q=0}$ is
nonvanishing, indicating spontaneous breakdown
of chiral symmetry. Secondly, the condensate is nonanalytic
in the coupling constant $G^2$, thus indicating
nonperturbative effects: although it seems natural from
dimensional analysis that $\langle 0 | \overline{\psi}\psi |0\rangle
\propto \sqrt{G^2}$ for small quark masses, it
is impossible to obtain such a behavior
in perturbation theory where one can only
generate terms $\propto G^{2n}$. Thirdly, one may suspect
that there is a phase transition in $QCD_{1+1}$.

Since the coupling constant $g$ in $QCD_{1+1}$ carries the dimension
of a mass, the theory is super-renormalizable and the
scale is set both by the coupling and by the mass.
In practice this implies that
$\langle 0 | \overline{\psi}\psi |0\rangle$
can (up to some dimensionful overall factor)
depend on $G^2$ only through the combination
$\alpha \equiv G^2/m_q^2$.
Therefore, in order to address the abovementioned issues
of nonperturbative effects and a possible phase
transition, it is necessary to consider nonzero quark masses.\\[2.ex]

\noindent
{\bf 2. $\langle 0|\bar{\psi}\psi|0\rangle$ from
sum rules}\\[1.5ex]
For nonzero quark masses, the vacuum expectation value
of the scalar density diverges already for free fields
\begin{equation}
\langle 0|\bar{\psi}\psi |0\rangle = \frac{N_C}{2\pi}
m_q\log \frac{\Lambda^2}{m_q^2}
{}.
\label{eq:freediv}
\end{equation}
However, due to the mild UV behavior in $1+1$ dimensions
($QCD_{1+1}$ is super-renormalizable) it is sufficient to
subtract the free field expectation value (i.e. to
``normal order'') to render
$\langle 0|\bar{\psi}\psi |0\rangle$ finite. This motivates
the definition
\begin{equation}
\left. \langle 0|\bar{\psi}\psi |0\rangle \right|_{ren}
\equiv
\langle 0|\bar{\psi}\psi |0\rangle -
\left. \langle 0|\bar{\psi}\psi |0\rangle \right|_{g=0}.
\label{eq:subtract}
\end{equation}
The condensate itself can be evaluated using current algebra
\begin{eqnarray}
0 &=& \lim_{q \rightarrow 0} iq^\mu \int d^2x e^{iqx}
\langle 0|T\left(\bar{\psi}\gamma_\mu \gamma_5\psi (x)
\bar{\psi}i\gamma_5\psi (0)\right) |0\rangle
\nonumber\\
&=&-\langle 0|\bar{\psi}\psi |0\rangle
-2m_q \int d^2x
\langle 0|T\left(\bar{\psi}i \gamma_5\psi (x)
\bar{\psi}i\gamma_5\psi (0)\right) |0\rangle .
\label{eq:cural}
\end{eqnarray}
Upon inserting a complete set of meson states\footnote{
Because we are working at leading order in $1/N_C$, the
sum over one meson states saturates the operator product
in Eq.(\ref{eq:cural}).}
one thus obtains
\begin{equation}
\langle 0|\bar{\psi}\psi |0\rangle
= -m_q \sum_n \frac{f_P^2(n)}{M_n^2},
\label{eq:naivesum}
\end{equation}
where $
f_P(n)\equiv \langle 0|\bar{\psi}i\gamma_5\psi |n\rangle
=\sqrt{\frac{N_C}{\pi}} \int_0^1 dx \frac{m_q/2}{x(1-x)} \phi_n(x)\ \
$ and the wavefunctions $\phi_n$ and invariant masses $M_n^2$ are
obtained from solving
`t Hooft's bound state equation for mesons in
$QCD_{1+1}$
\begin{equation}
M_n^2 \phi_n(x) = \frac{m_q^2}{x(1-x)}\phi_n(x) + G^2 \int_0^1 dy
\frac{\phi_n(x)-\phi_n(y)}{(x-y)^2}.
\label{eq:thooft}
\end{equation}
The variable $x$ corresponds to the light-front
momentum fraction carried by the quark in the meson.
Note that `t Hooft's equation was been derived using
light-front quantization --- we will return to this point below.

In the limit of highly excited mesons, the masses and coupling
constants scale:
$M_n^2 \stackrel{n\rightarrow \infty}{\longrightarrow}n\pi^2 G^2$,
$f_P(n) \stackrel{n\rightarrow \infty}{\longrightarrow}\sqrt{N_C \pi G^2}$
\cite{ei:qcd} and thus the sum in Eq.(\ref{eq:naivesum}) diverges
logarithmically. Of course this only reflects the free field divergence
(\ref{eq:freediv}). In order to regularize Eq.(\ref{eq:naivesum})
in a gauge invariant way we introduce an invariant mass cutoff
and obtain
\begin{equation}
\left.\langle 0|\bar{\psi}\psi |0\rangle \right|_{ren}
= -m_q \lim_{\Lambda \rightarrow \infty}
\left[\sum_n \frac{f_P^2(n)}{M_n^2\left(1+M_n^2/\Lambda^2\right)}
-``g=0''\right].
\label{eq:regsum}
\end{equation}
Eq.(\ref{eq:regsum}) can be used to calculate
$\left.\langle 0|\bar{\psi}\psi |0\rangle \right|_{ren}$
numerically with high precision.

In order to generate an exact result we will use a trick and
replace the sum in Eq.(\ref{eq:regsum}), where both small
and large $n$ contribute, by a sum which is dominated by
large $n$ only. Due to lack of space, only the basic ideas of
the derivation will be sketched here ---
a more detailed discussion of the limit $\Lambda \rightarrow
\infty$ can be found in Ref.\cite{mb:tbp}.
For this purpose, let us consider
\cite{ei:qcd}
\begin{equation}
G(x,\Lambda)\equiv \sqrt{\frac{N_C}{\pi}}
\sum_n \phi_n(x) \left( \frac{1}{M_n^2}-
\frac{1}{M_n^2+\Lambda^2}\right) f_P(n).
\label{eq:defgr}
\end{equation}
Obviously this ``Green's function'' can be used to
compute the condensate, via
\begin{equation}
\left.\langle 0|\bar{\psi}\psi |0\rangle \right|_{ren}
=-m_q^2 \lim_{\Lambda \rightarrow \infty}
\int_0^1 \frac{dx}{x}
\left[G(x,\Lambda)-\left.G(x,\Lambda)\right|_{g=0}\right].
\end{equation}
{}From the equation of motion (\ref{eq:thooft}) one can
show that $f_P(n)=\frac{M_n^2}{2m_q}\sqrt{
\frac{N_C}{\pi}}\int_0^1 dy \phi_n(y)$. If one furthermore
invokes completeness of `t Hooft's wavefunctions, i.e.
$\sum_n \phi_n(x)\phi_n(y) = \delta(x-y)$
one can simplify the first term in $G(x,\Lambda)$,
yielding
$\sqrt{\frac{N_C}{\pi}}\sum_n \phi_n(x)f_P(n)/M_n^2
=\frac{N_C}{\pi}\frac{1}{2m_q}$ --- independent
of $x$ and the coupling constant. This term thus
drops out completely when we subtract the free field
Green's function.

The crucial point is that, for $\Lambda^2\rightarrow
\infty$, the remaining term in $G(x,\Lambda)$
is dominated by the $n \rightarrow \infty$:
each individual meson yields a negligible
contribution $\propto (M_n^2+\Lambda^2)^{-1}$ when
we send the cutoff to infinity, and a nonzero result
arises only from the summation over infinitely many
highly excited meson states. We have thus succeeded
in converting the {\it low energy sum rule}
(\ref{eq:regsum}) into a
{\it high energy sum rule} and we can now make use
of the abovementioned scaling properties of meson masses $M_n^2$ and coupling
constants $f_P(n)$ as well as of the wavefunction itself
$\phi_n(x) \stackrel{n \rightarrow \infty}
{\longrightarrow} \Phi(M_n^2x)$. The scaled
wavefunction
$\Phi(z) \equiv \lim_{n\rightarrow \infty}
\phi_n(z/M_n^2)$ satisfies the integral equation \cite{ei:qcd}
\begin{equation}
\Phi(z) = \frac{m_q^2}{z}\Phi(z) + G^2 \int_0^\infty dy
\frac{\Phi(z)-\Phi(y)}{(x-y)^2}
\label{eq:scaled}.
\end{equation}
In terms of these scaled quantities we thus find
\begin{equation}
\left.\langle 0|\bar{\psi}\psi |0\rangle \right|_{ren}
= \frac{N_C}{\pi}m_q^2 \int_0^1 \frac{dx}{x}
\left[ \sqrt{G^2}\sum_{n=1,3,..} \frac{\Phi(n\pi^2G^2x)}{n\pi^2G^2+\Lambda^2}
-``g=0'' \right]
\end{equation}
Upon performing the substitution $z=n\pi^2G^2 x$,
replacing
$\sum_{n=1,3,..} \rightarrow (2\pi^2G^2x)^{-1}\int_0^\infty dz$
\footnote{This is exact for $\Lambda \rightarrow \infty$
since the series receives nonvanishing contributions
only from the $n \rightarrow \infty$ region.}
and performing the $x$-integral one ends up with
\begin{equation}
\left.\langle 0|\bar{\psi}\psi |0\rangle \right|_{ren}
= \frac{N_C}{\pi}m_q^2 \left[ \frac{1}{\sqrt{G^2}}
\int_0^\infty \frac{dz \log z}{z} \Phi(z)
-``g=0'' \right].
\end{equation}
Note that $\frac{1}{\sqrt{G^2}}
\int_0^\infty \frac{dz}{z} \Phi(z)=\frac{\pi}{m_q}$ \cite{br:scal}
is independent of $G^2$ and it does not matter that the argument
of the logarithm is dimensionful, since a free theory
subtraction is performed.
While `t Hooft's equation cannot been solved
exactly, the scaling equation can be solved
analytically \cite{br:scal}. The lengthy expression
for $\Phi(z)$ will not be displayed here and we
merely give the result for the renormalized
condensate
\begin{equation}
\left.\langle 0|\bar{\psi}\psi |0\rangle \right|_{ren}
= \frac{m_qN_C}{2\pi}
\left\{ \log \left(\pi \alpha\right)  -1 -\gamma_E+
\left(1-\frac{1}{\alpha}\right)
\left[ (1-\alpha) I(\alpha)-\log 4\right]\right\},
\end{equation}
where $\alpha =G^2/m_q^2$, $\gamma_E =.5772..$ is Euler's
constant and
\begin{equation}
I(\alpha) = \int_0^\infty \frac{dy}{y^2}
\frac{ 1 - \frac{y}{\sinh y \cosh y}}
{\left[\alpha (y \coth y-1)+1\right]}.
\label{eq:exact}
\end{equation}
This result is exact for $N_C\rightarrow \infty$ and
{\it all} quark masses. In the limit $\alpha \rightarrow
\infty$ one recovers Zhitnitsky's result Eq.(\ref{eq:mis0}).
Furthermore, one can verify that
the exact result coincides with the numerical
evaluation of Eq.(\ref{eq:regsum}). In the limit
$\alpha \rightarrow 0$ the condensate vanishes, which
is not surprising since we have subtracted the free
field result.
\begin{figure}
\begin{Large}
\unitlength1.cm
\begin{picture}(14,7.5)(2.3,-10.4)
\put(15.8,-10.3){\makebox(0,0){$m_q/\sqrt{G^2}$}}
\put(3.4,-4.2){\makebox(0,0){$\frac{
\langle 0|\bar{\psi}\psi|0\rangle}{N_C\sqrt{G^2}}$}}
%10= lange bild , 5=hohe bild
%0,12 veschiebung bzgl. linke untere Ecke
%alles in Einheiten von unitlength
\includegraphics{vac.ps}
\end{picture}
\end{Large}
\caption{Renormalized quark condensate as a function of the
quark mass. Both in units of the effective coupling
$G^2=g^2C_F/\pi$. Note the absence of singularities.}
\end{figure}

\noindent
{\bf 3. Discussion}\\[1.5ex]
We have started from `t Hooft's equation which is based
on light-front quantization. The light-front vacuum
is trivial, i.e. identical to the Fock space vacuum \cite{le:ap}.
Nevertheless, using current algebra and sum rule
techniques, we obtained a nonzero result for the
quark condensates. The result we obtained agrees with
numerical calculations using equal time quantization
(see Refs.\cite{le:ap,li:vac} for $m_q=0$ and
Ref.\cite{mth:priv} for the general case).

The exact result which we have obtained (\ref{eq:exact})
is an analytic function of $\alpha$ in the complex
plane cut along the negative real axis --- i.e. there
is {\it no} phase transition. An asymptotic expansion
in powers of $\alpha$ yields
\begin{equation}
\frac{2\pi}{m_q N_C} \left.\langle 0|\bar{\psi}\psi |0\rangle \right|_{ren} =
\sum_{\nu=1}^\infty c_\nu \alpha^\nu,
\label{eq:asym}
\end{equation}
where the coefficients show factorial growth
\begin{equation}
c_\nu \stackrel{\nu \rightarrow \infty}{\sim}
(-1)^\nu e^{-2} 2^{1-\nu}(\nu-1)!\quad,
\end{equation}
i.e., the asymptotic expansion for
$\left.\langle 0|\bar{\psi}\psi |0\rangle \right|_{ren}$
is only Borel summable and the Borel series has a
finite radius of convergence.
Applying the inverse Borel transfer to the Borel
summed series one recovers the exact result which
reflects the absence of terms like $e^{-\frac{1}{\alpha}}$.
I have compared the first three terms in the
asymptotic expansion with the perturbative (Feynman diagrams) expansion
and found agreement.
Nevertheless Eq.(\ref{eq:exact}) is a completely
nonperturbative result,
because one has to sum up {\it all} terms in the
perturbation series before one obtains the right
scaling behavior (\ref{eq:mis0}) for small quark mass
(large $\alpha$). It is also {\it not} sufficient to
keep only the asymptotic behavior of the series: it is easy to write
down an expression which has the same asymptotic
coefficients for large $\nu$ but does not yield the desired
$\sqrt{\alpha}$ behavior for $m_q\rightarrow 0$.

\end{document}